\documentclass[%
 reprint,
superscriptaddress,
 amsmath,amssymb,
 aps,
floatfix,longbibliography
]{revtex4-2}

\usepackage{graphicx}
\usepackage{dcolumn}
\usepackage{bm}
\usepackage{siunitx}
\usepackage{float}
\usepackage{natbib}
\usepackage{appendix}
\usepackage{ulem}
\usepackage{comment}
\usepackage[pagewise]{lineno}

\begin{document}

\preprint{APS/123-QED}
    
\title{Geometric frustration and pairing-order transition in confined bacterial vortices} 

\author{Kazusa Beppu}\email{kazusa.beppu@aalto.fi}
\affiliation{Department of Physics, Kyushu University, Motooka 744, Fukuoka 819-0395, Japan}\affiliation{Department of Applied Physics, Aalto University School of Science, Puumiehenkuja 2, Espoo, 02150, Finland}


\author{Kaito Matsuura}
\affiliation{Department of Physics, Kyushu University, Motooka 744, Fukuoka 819-0395, Japan}

\author{Yusuke T. Maeda}\email{maeda@cheme.kyoto-u.ac.jp}\affiliation{Department of Physics, Kyushu University, Motooka 744, Fukuoka 819-0395, Japan}\affiliation{Department of Chemical Engineering, Kyoto University, Nishikyo-ku, Kyoto 615-8246, Japan}


\date{\today}

\begin{abstract}
Dense systems of active matter exhibit highly dynamic collective motion characterized by intermingled vortices, referred to as active turbulence. The interaction between these vortices is key to controlling turbulent dynamics, and a promising approach for revealing the rules governing their interaction is geometric confinement. In this study, we investigate the vortex-pairing patterns in confined bacterial suspensions as a model frustrated system in which a perfect antiferromagnetic state is prohibited. We found that three-body vortex interactions exhibited anomalous pairing-order transition from co-rotational vortex-pairing to counter-rotating patterns with frustration. Although an active matter system is in non-equilibrium, our theory based on bending energy accounts for significant features including pattern transition and shift of the transition point in frustrated systems. Moreover, the interplay between the chirality in collective motion and frustration in vortex pairing creates a collective rotational flow under the broad geometric conditions of a confined space. Our results show that frustrated vortex patterning promotes a geometric approach for arranging active turbulence in microfluidic systems.
\end{abstract}

\maketitle

\section{Introduction}
Autonomous biological systems such as molecular motor proteins \cite{needleman2017active} convert chemical energy into mechanical action. The exploration of such autonomous motion, stemming from out-of-equilibrium dynamics, poses a significant challenge in science, particularly in physics and biology. A particular system, termed active matter \cite{ramaswamy2010mechanics,marchetti2013hydrodynamics}, is a class of materials such as motor proteins, swimming bacteria, eukaryotic cells, and even animals, that can move spontaneously. The emergent dynamics of active matter are collectively aligned motions at high densities, fostering self-organized collective motion. Such collective motions are not limited to biological systems, and similar cluster and lane formations are also found in non-biological matters, such as active colloids \cite{bechinger2016active,zottl2016emergent,elgeti2015physics} and swimming droplets \cite{izri2014self,maass2016swimming}. The concept of active matter has recently advanced into micro-robotics engineering; therefore, controlling their collective motion can provide a deeper understanding of the mass and energy transport found in nature and technology.

Active turbulence, which is characterized by the emergence of disordered structures akin to turbulence, represents the foundational collective motion within dense active matter \cite{alert2022active,aranson2022bacterial}. The active turbulence occurs at a microscopic scale (i.e., at low Reynolds numbers), and the spontaneous stirring of viscous fluids offers practical utility in their manipulation within microfluidic systems. Among the microscopic systems employed to investigate active turbulence, swimming bacteria are a widely used model \cite{wensink2012meso,dunkel2013fluid,lushi2014fluid,peng2021imaging,chen2021confinement}. When a suspension of bacteria is confined within a water-in-oil droplet whose size matches the velocity-correlation length, a single vortex-like rotation stabilizes within the droplet \cite{wioland2013confinement,hamby2018swimming}. Furthermore, the geometrical boundaries that make vortices interact with neighbors, such as lattice cavities connected via channels \cite{wioland2016ferromagnetic}, multiplet lattice-free microwells \cite{beppu2017geometry}, and pillar spacing \cite{nishiguchi2018engineering}, enables control of the pairing order of vortices in the same or opposite directions, depending on geometric parameters such as channel width and spacing distance. These different patterns of rotating interacting vortices are called ferromagnetic vortex patterns (FMV) and antiferromagnetic vortex patterns (AFMV), analogous to the Ising model system in magnetism. This concept was introduced by Wioland et al. \cite{wioland2016ferromagnetic}, which allowed us to explore the connection between equilibrium spin systems and active matter systems. Furthermore, a numerical study of active turbulence with periodic obstacles revealed a continuous second-order phase transition behavior from static vortex lattice to active turbulence by varying the advection strength as an effective temperature in the argument of the Ising model system \cite{reinken2022ising}. The study of self-organized patterns under geometrical constraints has been applied not only to bacterial active turbulence but also to various systems such as nematic cell populations \cite{deforet2014emergence,duclos2017topological,duclos2018spontaneous,guillamat2022integer,ienaga2023geometric}, active cytoskeletons driven by motor proteins \cite{wu2017transition,fan2021effects,opathalage2019self,hardouin2019reconfigurable,araki2021controlling,hardouin2022active}, and active colloidal particles \cite{bricard2015emergent,van2016spatiotemporal,bartolo2021topology}, thus advancing an understanding of common rules in active matter systems.

Despite intense research efforts on active turbulence under geometric constraints, the effect of frustration on vortex rotations, observed when three vortices interact anti-ferromagnetically, remains unclear. Dense bacterial suspensions confined in microwells with triangular lattices reportedly exhibit highly ordered ferromagnetic states \cite{wioland2016ferromagnetic}. In active nematics, a dynamic change in the collective flow in the direction of rotation was found in a mixture of microtubules and kinesins confined in a triplet annular channel owing to the accumulation of frustration \cite{hardouin2020active}. Recent theoretical studies have also demonstrated that polar active fluids exhibit topologically protected sound modes in the presence of periodic structures with Kagome lattice geometry \cite{sone2019anomalous}, and that the symmetry breaking of turbulent vortices in active turbulence occurs to avoid frustration \cite{nishiguchi2018engineering,reinken2020organizing}. Moreover, frustrated interactions among self-propelled particles can result in clogging and jamming \cite{lei2023collective,sevc2012geometrical,reichhardt2021active}. As such, geometric frustration plays an important role in pattern formations. Therefore, elucidating the relationship between geometric frustration and emergent order is essential for understanding the physical mechanism controlling active turbulence.

This study revealed the effect of frustration on the vortex pairings of confined bacterial suspensions using lattice-free geometric confinement, thus enabling the intrinsic nature of interacting vortices intertwined with geometrical boundaries to be elucidated. These suspensions exhibited two distinctive patterns for three vortex pairings. One pattern entails all three vortices rotating uniformly in a singular direction, and the other pattern represents a frustrated state, characterized by one vortex rotating in the opposite direction to the others. A simple theoretical model based on bending elasticity in orientation can elucidate the shift in the transition point due to the cost in the bending distortion of the polar orientation field, suggesting that not only the distance between vortices but also their geometrical location is important for understanding the rule of vortex pairing. Moreover, the interplay between the chiral vortex and geometric frustration allows us to control active turbulence to a larger rotational flow than either effect alone. Our findings advance the understanding of geometry-based design principles for controlling active fluids.

\begin{figure*}[htb]
\begin{center}
\includegraphics[width=16cm]{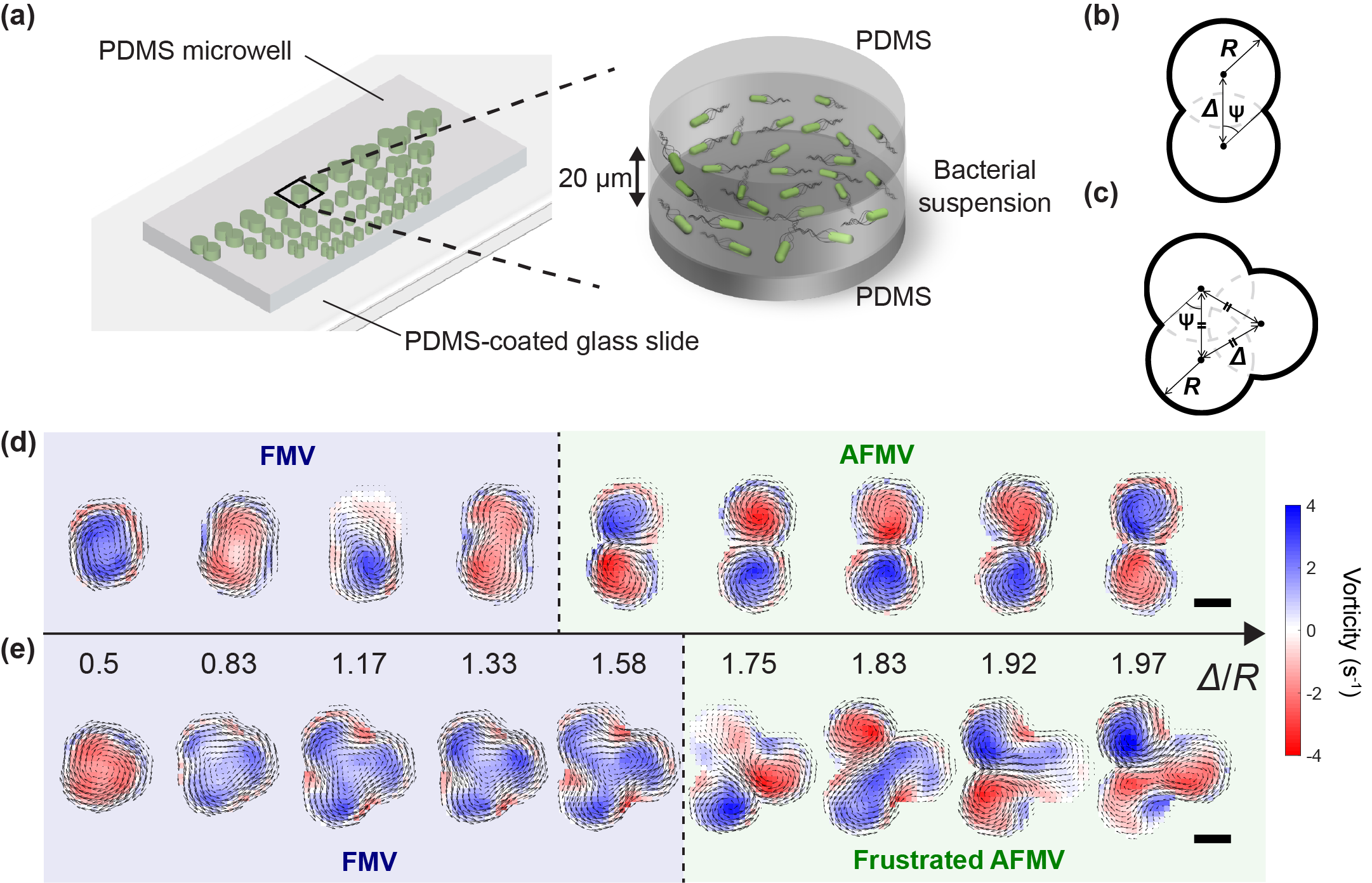}
 \caption{\textbf{Frustrated bacterial vortices confined in the designed microwells.} (a) Schematic of a polydimethylsiloxane (PDMS) microdevice enclosing dense bacterial suspensions. (b, c) Design of overlapping circular boundary. The doublet or triplet circular boundary is defined with the circle radius $R$ and the distance between the center of the circle $\Delta$. The angle $\Psi$ is defined as $\cos\Psi = \frac{\Delta}{2R}$. (d, e) Bacterial vortices under triplet circle confinement. The bacterial suspension is enclosed in a PDMS microwell with PDMS top surface. The presented vorticity map of collective motion is obtained by averaging the velocity fields for 10 s for microwells with a radius of $R=$\SI{19}{\micro\meter}. Scale bars: \SI{20}{\micro\meter}.}\label{fig1}
\end{center}\end{figure*}

\section{Materials and Methods}
\subsection{Bacterial culture}
\textit{Escherichia coli} (strain RP4979) were cultured overnight by inoculating a single bacterial colony into \SI{1}{\milli\liter} of Lysogeny broth medium (1\% tryptone, 1\% NaCl, 0.5\% yeast extract) and incubating it on a rotary shaker inside an incubator (150 rpm) at \SI{37}{\celsius}. The saturated culture was diluted 200 fold in a tryptone broth medium (1\% tryptone, 1\% NaCl) and further incubated on a rotary shaker in an incubator at \SI{30}{\celsius} at 150 rpm for 5 h until the optical density at \SI{600}{\nano\meter} of the culture reached approximately $0.4$. To obtain dense bacterial suspensions that formed active turbulence, we centrifuged the cultured suspension at 3000 rpm at \SI{25}{\celsius} for 10 min. The final body volume fraction of bacteria was approximately $20\% v/v$.  

\subsection{Microfabrication and sample preparation}
Dense bacterial suspensions undergoing bacterial turbulence were enclosed in polydimethylsiloxane (PDMS) microwells with a height of \SI{20}{\micro\meter} (Fig. \ref{fig1}(a)), which were fabricated using standard soft lithography, following previously described protocol \cite{beppu2017geometry,beppu2021edge}. In this study, two different surface conditions were used to confine bacteria. The first is for an achiral bacterial vortex, in which the entire surface is composed of PDMS coated with a nonionic surfactant solution (Triton X-100, Sigma-Aldrich) (symmetric chamber) \cite{beppu2017geometry}. The surface treatment was done by immersing the PDMS microwells in a 10\% Triton X-100 solution for 3 h and then rinsing them with water to prevent the substrate from absorbing bacteria. Then, the dense bacterial suspensions were encapsulated by gently sandwiching them between a PDMS-coated glass slide and the fabricated PDMS microwells. The second surface condition is for a chiral bacterial vortex, in which only the top surface is replaced with a water-in-oil interface (asymmetric chamber) to induce vortex rotation via the surface chiral swimming of bacteria \cite{beppu2021edge}. The sample with the asymmetric chamber was created as follows. A PDMS thin film with patterned microwells was adhered to a glass slide and treated by immersion in a solution of polyethylene glycol-poly-L-lysine (Nanocs, PG2k-PLY) immediately after exposure to air plasma for 1 min (plasma cleaner, Harrick Plasma). After a flow cell was assembled with two parallel double-sided tapes of \SI{100}{\micro\meter} thickness and a coverslip, dense bacterial suspensions were gently loaded into the flow cell, and the microwells were sealed by injecting oil (light mineral oil, Sigma Aldrich) with a surfactant (SPAN80, Nacalai) at 2.0wt\%, and the excess suspensions over the microwells were flushed out and absorbed with filter paper from the other side of the flow cell. Both ends were sealed with epoxy glue (Huntsman Ltd.) to prevent residual flow.

\subsection{Microscopy and data processing}
Bright-field microscopy was used to image bacterial suspensions using an inverted microscope (IX73, Olympus) with a $20 \times$ objective lens. Videos were recorded using a charge-coupled device camera (DMK23G445, Imaging Source) at 30 fps. The velocity field of the bacterial collective motion was measured via conventional particle image velocimetry (PIV). using the PIVLab toolbox in MATLAB. Images were preprocessed using the Wiener2 denoise filter, and the interrogation window size was set to $16 \times 16$ pixels$^2$ ($2.98 \times \SI{2.98}{\micro\meter^2}$), which is comparable to the typical body length of bacteria. The subsequent velocity fields were smoothed by averaging over $\SI{1}{\second}$, which corresponded to the typical lifetime of the turbulent vortices in each time frame.

\section{Results}

The designed boundary of the microwells consists of overlapping circles. When the boundary is a symmetrical arrangement of two (or three) overlapping circles of radius $R$ equally spaced by a distance $\Delta$ between the centers of the circles, this is referred to as a doublet (or triplet) circle pattern (Fig. \ref{fig1}(b) and (c)). Confined bacterial vortices can mutually interact within a microwell.

 For a doublet circular boundary with $R$=\SI{19}{\micro\meter} (Fig. \ref{fig1}(b)), a co-rotational pairing pattern of FMV (two vortices rotate in the same direction) appears at small $\Delta/R$, but an anti-rotational pairing pattern of AFMV (two vortices rotate in the opposite direction) emerges at $\Delta/R \geq 1.33$ (Fig. \ref{fig1}(d) and Movie 1). This is consistent with previous findings that the transition between FMV and AFMV occurs at $\Delta_c/R = \sqrt{2}$ \cite{beppu2017geometry,beppu2021edge}. This rule is derived from the geometry-dependent effective energy of the polar interactions of the particles at the tip where two circles intersect.

\begin{figure*}[tb]
\begin{center}
\includegraphics[width=15cm]{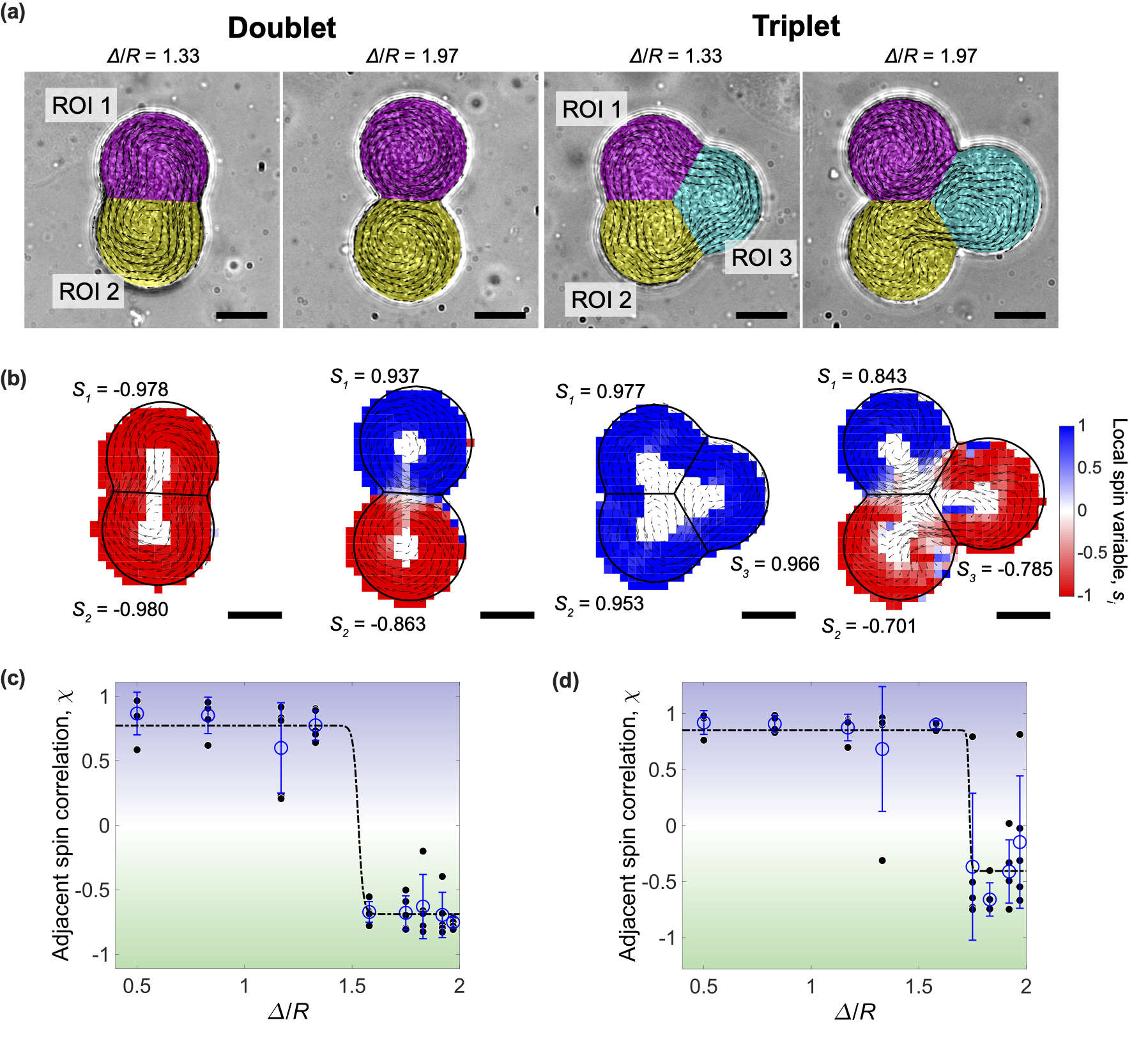}
 \caption{\textbf{Vortex-pairing patterns in a confined bacterial suspension}. (a) Microscopic images and region of interest (ROI) settings. Doublet circular boundary (left) and triplet circular boundary (right) have two and three ROIs, respectively. The velocity maps are analyzed in these ROIs. (b) Maps of local spin variables $s_i$ calculated within \SI{15}{\micro\meter} apart from the boundary. The instantaneous velocity fields and corresponding spin variables are displayed in (a) and (b). Scale bars: \SI{20}{\micro\meter}. The spin variable $S_i$ averaged within each ROI ($i=1, 2, 3$) is shown with its number. (c) Geometry-induced transition of bacterial vortex pairings. The spin correlation function $\chi$ is obtained from the bacterial vortex and plotted as a time average for 10 s. For a doublet case, if $\chi$ is positive and close to +1, it is an FMV pattern, and if negative and close to $-0.5$, it is an AFMV pattern. The doublet case shows an abrupt change at $\Delta/R$=1.4. (d) Geometry-induced transition of frustrated AFMV pattern. If bacterial vortices show a frustrated AFMV pattern, $\chi$ is negative and close to $-0.5$. The triplet case shows an abrupt change at approximately $\Delta/R$=1.7. The dotted line is fitted with a sigmoidal function. The black dots represent the individual experimental data, and the blue circles represent their mean values. The error bar is the standard deviation.}\label{fig2}
 \end{center}
 \end{figure*}
 
 For three vortices (Fig. \ref{fig1}(c)), an FMV pattern in which all vortices have the same orientation can be one of the stable states. However, if one vortex has the opposite orientation, geometric frustration occurs such that the AFMV and FMV patterns coexist. As the shift in the transition point from the FMV to AFMV pattern in frustrated vortices has not been elucidated, we experimentally investigated the geometric conditions under which the frustrated AFMV pattern appeared. The bacteria inside the triplet circular boundary formed an FMV pattern of three interacting vortices with small geometric parameters $\Delta/R \leq 1.58$. When $\Delta/R$ was further increased to $\Delta/R \geq 1.75$, one of the three vortices showed counter-rotation, and the AFMV pattern coexisted with the FMV pattern (Fig. \ref{fig1}(e) and Movie 1). In some cases, the three vortices were split into several small vortices such that the pattern would not be frustrated (Fig. S1 in the Supplemental Material \cite{supplement}). This suggests that geometric frustration can yield nontrivial patterns that are not observed in frustration-free systems.

We analyzed a spin variable, $s_i$, defined by the velocity component within the $i$-th region of interest (ROI) to quantify the order of the emergent patterns (Fig. \ref{fig2}(a)). 
\begin{equation}
s_i = \frac{\mathbf{v}(\mathbf{r},t)\cdot \mathbf{t}}{|\mathbf{v}(\mathbf{r},t)|} 
\end{equation}
where $\mathbf{v}(\mathbf{r},t)$ is the velocity field obtained from PIV analysis as a function of position $\mathbf{r}$ and time $t$, and $\mathbf{t}$ denotes the tangential unit vector along the boundary. The ROI was set to be within \SI{15}{\micro\meter} of the boundary of each circle. Its ensemble average within the ROI, $S_i = \langle s_i \rangle_{\mathbf{r}\in \textrm{ROI}_i}$, indicates whether the vortices rotate clockwise (negative sign) or counterclockwise (positive sign) in the two or three circles. The spatiotemporal pattern is quantified as ``spin'' in terms of the vortex-pairing pattern (Fig. \ref{fig2}(b)). When the values of $S_i$ are mapped near the doublet/triplet circular boundaries, $S_i$ values of the same signs are occupied in the FMV pattern, whereas $S_i$ values of opposite signs coexist in the frustrated AFMV pattern (Fig. \ref{fig2}(b)).

\begin{figure*}[tb]
\begin{center}
\includegraphics[width=14cm]{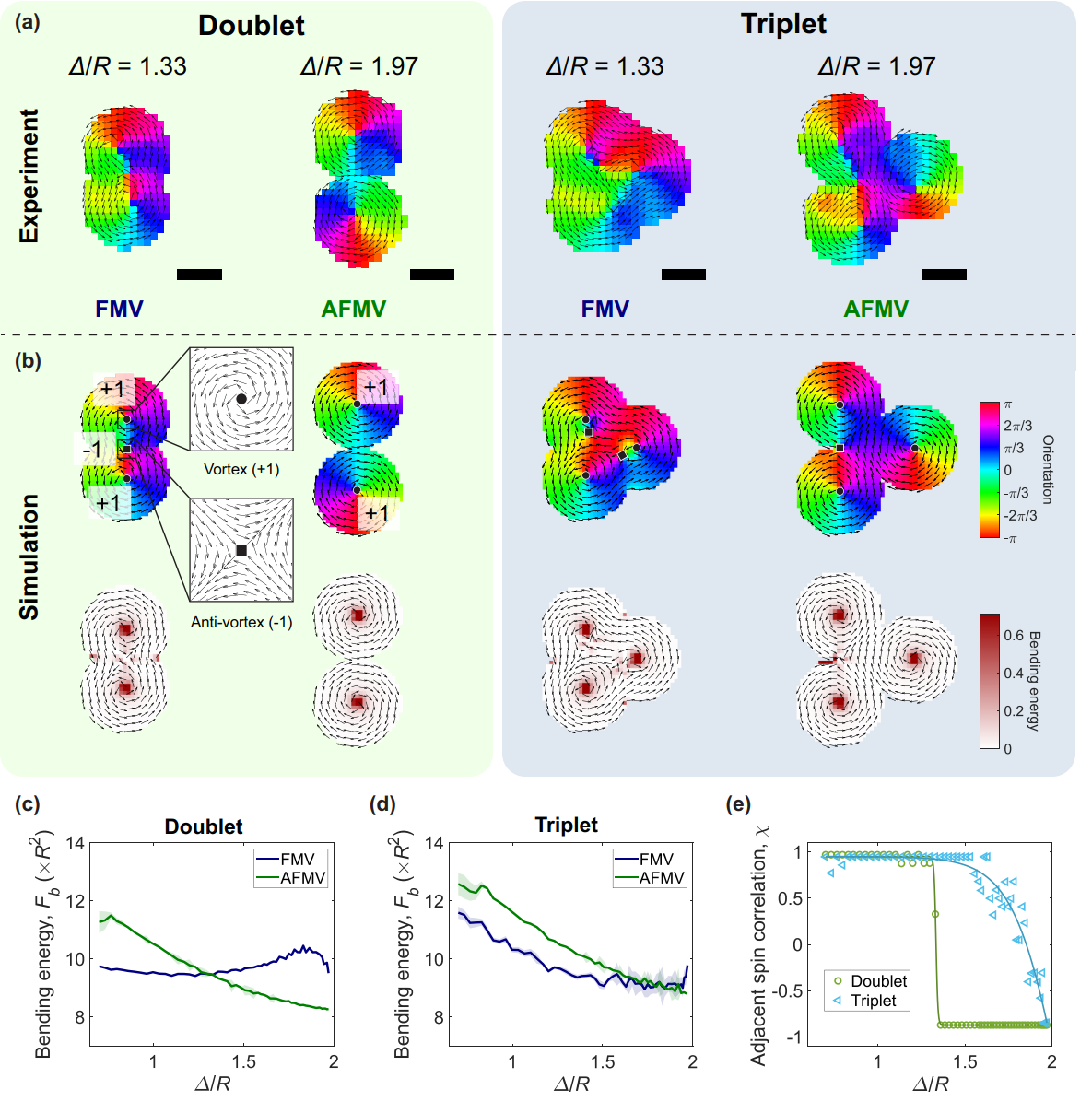}
 \caption{\textbf{Theoretical analysis of the transition of bacterial vortex pairings.} (a) Representative snapshots of the velocity orientation fields in the doublet (left) and triplet (right) circular boundaries. Scale bars: \SI{20}{\micro\meter}. (b) Representatives of numerical simulations corresponding to (a). The upper and lower part of the figure denote the orientation field and the local bending energy $(\bm{P}\times(\bm{\nabla}\times\bm{P}))^2$, respectively. The black circles and squares indicate the locations of topological defects with $\pm1$ charges, respectively. (c, d) Geometry-dependent bending energy density for numerically obtained FMV and AFMV patterns in the doublet (left) and triplet (right) circular boundaries. The spatially averaged bending energy $F_b$ is further scaled by $R^2$ for comparison with experiments. The colored areas represent the standard deviations of the final states for twenty simulations evolved from different random initial states. The standard deviation is large, particularly for the FMV patterns at the triplet circular boundary of large $\Delta/R$, because the location of the emerging two anti-vortices tends to vary from simulation to simulation. (e) Geometry-dependent adjacent spin correlation for numerically obtained FMV and AFMV patterns in the doublet and triple circular boundaries. The spin correlation was calculated using Eq. (2), where the ROI was set to be within 0.8$R$ of the boundary. The final states of 20 simulations of FMV and AFMV, respectively, were considered together, and the spin correlations associated with the bending energies from the smallest to the 20th smallest energies were averaged. The lines indicate fitting curves with sigmoidal functions.}
\label{fig3}
 \end{center}
 \end{figure*}

This study primarily investigate the occurrence of pairing-order transition in the interactions of bacterial spins representing vortices. This investigation requires an evaluation of whether adjacent spins are oriented in the same direction. To this end, we define the spin correlation function $\chi$ as 

\begin{equation}
\chi = \textrm{sign}(\langle S_i \cdot S_j \rangle_{i,j})\langle |S_i \cdot S_j| \rangle_{i,j}
\end{equation} \label{eq2}
where $\langle \cdot \rangle_{i,j}$ is the ensemble average for all combinations of spin variables, and $\textrm{sign}(\cdot)$ is the sign function, which gives a positive or negative sign for the multiplication of the spin variable. If one of the two or three vortices has opposite spin directions, the sign function is negative. This results in $\chi=1$ for the FMV pattern and $\chi \simeq -0.8$ for the AFMV pattern (Fig. \ref{fig2}(c)). Using this change in parameter as an order parameter, we examined the geometric dependence of the pattern transition.

For a doublet circular boundary with a confinement size of $R=\SI{19}{\micro\meter}$, which is comparable to the typical vortex size in bacterial active turbulence, the $\chi$ parameter is close to $+ 0.8$ at $\Delta/R \leq 1.33$ but becomes negative and close to $-0.7$ at $\Delta/R \geq 1.58$ (Fig. \ref{fig2}(c)). However, in the triplet circular boundary, the $\chi$ parameter takes positive values even at $1.33 \leq \Delta/R < 1.75$ and becomes negative at $\Delta/R \geq 1.75$ (Fig. \ref{fig2}(d)). In the triplet circular boundary, the FMV pattern was favored compared with the doublet pattern, probably to avoid frustration. Moreover, we examine the temporal changes in both the spin correlation $\chi$ and spin variable $S_i$ in each ROI (Fig. S2 in the Supplemental Material \cite{supplement}). At the doublet circular boundary, both FMV and AFMV patterns are highly stable (Fig. S2(a)), with vortices in each ROI persisting throughout the observation (30 s) (Fig. S2(c)). By contrast, the order of the frustrated AFMV patterns at the triplet circular boundary tends to vary relatively widely (Fig. S2(b)). This stems from the large fluctuations in the spin variable of one of three interacting vortices, which are also characterized by the wide probability distributions (Fig. S2(d)). Particularly, vortices with large spin fluctuations tend to belong to the frustrated vortex pair, i.e., FMV pair, of the three vortex pair combinations in the frustrated AFMV pattern, implying that the accumulation of frustration of interacting vortices causes the large fluctuations of the frustrated AFMV pattern.

\begin{figure*}[tb]
\begin{center}
\includegraphics[width=14cm]{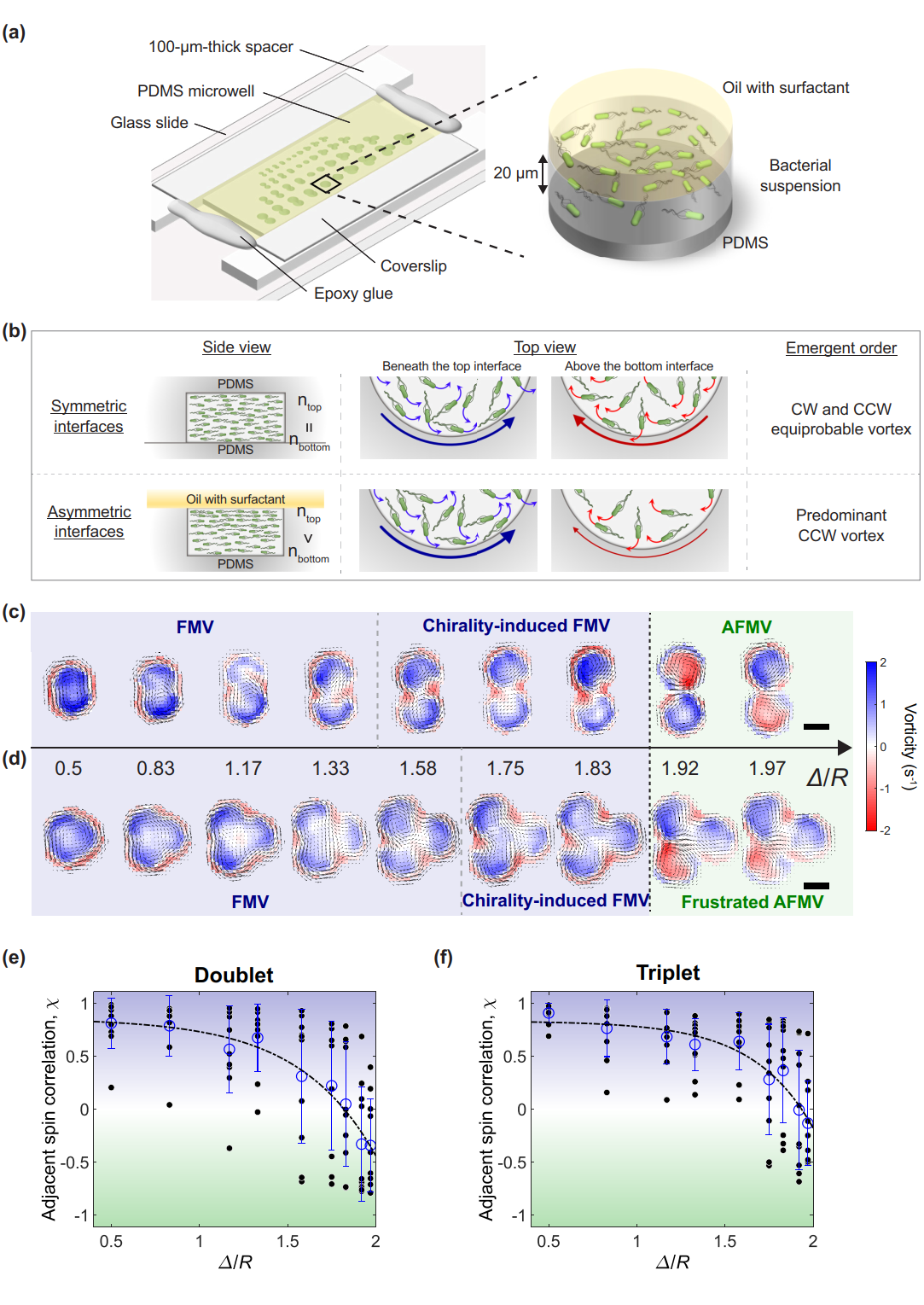}
 \caption{\textbf{Altering transition point of frustrated vortex pairing by chiral collective motion.}  (a) Schematic of a PDMS microdevice with a water-in-oil interface under the top of a microwell. (b) Schematic illustration of the mechanism of chiral vortex formation. The microwell with the asymmetric interfaces induces a higher density of bacteria at the top interface than the one at the bottom substrate ($\mathrm{n_{top}} > \mathrm{n_{bottom}}$) \cite{beppu2021edge}. The interaction between the interface-induced chiral motion of bacteria and the lateral boundary results in biased motion in clockwise and counterclockwise directions above and below the bottom and top interface, respectively, as viewed from above. Their competition leads to the CCW-predominant chiral vortex in the asymmetric system. (c, d) Chiral bacterial vortices under confinement with (c) doublet and (d) triplet circular boundary. The bacterial suspension was enclosed in a PDMS microwell with an oil--water top surface. The presented vorticity map of collective motion is for microwells with a radius size of $R=$\SI{19}{\micro\meter}. Scale bars: \SI{20}{\micro\meter}. (e, f) Geometry-induced transition of chiral FMV to (frustrated) AFMV pattern in (e) doublet and (f) triplet circular boundary. The blue circles represent mean values, and the error bar is the standard deviation.}\label{fig4}
 \end{center}
 \end{figure*}

Further, we sought a theoretical explanation of the shift in the transition point from FMV to AFMV in frustrated bacterial vortices. Polar active turbulence has been extensively investigated both experimentally and theoretically. The commonly used model, termed the Toner--Tu--Swift--Hohenberg (TTSH) equation \cite{wensink2012meso,dunkel2013fluid, grossmann2014vortex, reinken2018derivation, reinken2019anisotropic, puggioni2022giant}, has captured the statistical properties, such as energy cascades, of active polar suspensions in the bulk, and the effect of the geometrical constraints of pillar spacing on turbulent behavior \cite{nishiguchi2018engineering, reinken2020organizing}. Therefore, the TTSH equation is the basis for the quantitative description of active polar turbulence. However, the realization of vortex-pairing patterns in confined active turbulence remains under discussion \cite{shiratani2023route}. In the past study \cite{beppu2017geometry, beppu2021edge}, it was demonstrated that polar interactions around the tip of a doublet circular boundary determine emergent patterns, implying that orientation distortions are essential in pattern transition. This notion motivated us to consider the bending free energies of patterns that imitate FMV and AFMV patterns by focusing on the topological defects with $\pm1$ charges. A topological defect is a singularity that cannot be defined by the director and has a charge determined by the degree of rotation of the director field relative to the rotation axis as it circles the defect along the closed loop. The $\pm1$ charges correspond to $\pm2\pi$ rotations, respectively. As shown in the representative steady-state patterns of FMV and AFMV in Fig. \ref{fig3}(a), the FMV pattern in the doublet circular boundary includes two $+1$ topological defects at each center of circle, i.e., two vortices rotating in the same direction, and one $-1$ topological defect between them (an anti-vortex), whereas the AFMV pattern has two $+1$ topological defects with opposite rotational direction (Fig. \ref{fig3}(b)). For the triplet circular boundary, the FMV typically consists of three vortices rotating in the same direction in each circle and two anti-vortices between two of them, whereas the frustrated AFMV includes three vortices, one of which has an opposite rotation, and one anti-vortex (Fig. \ref{fig3}(b)). In the argument of the free energy of active polar materials \cite{marchetti2013hydrodynamics}, the presence of the topological defects is most penalized by the diffusion of alignment among active units. Here, we simulate the continuum model of the polarization field $\bm{P}(\bm{r}) = (\cos\theta, \sin\theta)$, where $\theta(\bm{r})$ denotes the angle at position $\bm{r}$ within a confined space (see the Supplemental Material for calculation details \cite{supplement}). In this model, the polarization field is evolved by the diffusion of alignment while undergoing perturbations in the presence of fixed topological defects, allowing us to reproduce both FMV and AFMV patterns at varying $\Delta/R$ (Fig. \ref{fig3}(b) and Fig. S3).

We calculated the bending energy densities of the FMV and AFMV patterns using $F_b = \langle ( \bm{P}\times(\bm{\nabla}\times\bm{P}))^2 \rangle_{\bm{r}}$, where $\langle \cdot \rangle_{\bm{r}}$ denotes the ensemble average over the confined space, by referring to the bending free energy for polarization fields in active polar fluid systems \cite{marchetti2013hydrodynamics}, inspired by the classical Frank free energy of nematic liquid crystals. Note that the other free energies (i.e., splay and twist) are neglected. Twist deformations can simply be ignored because the system is quasi-two-dimensional, while splay deformations are known to be important even in quasi-two-dimensional systems, e.g., for the dynamic patterns observed in a dense bacterial system \cite{liu2021viscoelastic}. However, due to the high concentration in our experimental system, it is reasonable to consider the polarization field divergence-free, given that the confined bacteria are uniformly distributed in a steady state, which is supported by the flow patterns such as the FMV and AFMV observed in the experiment. The difference in $F_b$ between the two patterns indicates whether a transition point exists where a stable pattern with low energy can be switched.

The bending orientation energies of the FMV and AFMV patterns in the doublet circular boundary are compared under different geometric conditions. When $\Delta/R$ is small and the overlap between the two circular boundary shapes is considerable, the FMV pattern has a lower bending energy than the AFMV pattern (Fig. \ref{fig3}(c)). This implies that more energy is required to form an AFMV pattern at small $\Delta/R$ and induce orientation bending. A reversal of the relationship between the FMV and AFMV patterns in terms of bending energy is observed at $\Delta/R >1.3$. 

Similarly, given the boundary condition of a triplet circular boundary, we calculated the bending elastic energy required to form the FMV and frustrated AFMV pattern orientations (Fig. \ref{fig3}(d)). The bending energy of the FMV pattern is large at small $\Delta/R$, compared to that of the doublet circular boundary, and decreases with increasing $\Delta/R$. This is because, at small $\Delta/R$, the three $+1$ topological defects and two $-1$ defects in between are close to each other, causing large bending energy due to large distortion of the polarization field, whereas, as $\Delta/R$ increases, the distortion of the polarization field is mitigated to some extent. By contrast, the bending energy of the frustrated AFMV shows the same decreasing trend as that of the AFMV at the doublet circular boundary but is increased by frustration. The bending energy relationship between FMV and frustrated AFMV patterns is reversed at $\Delta/R = 1.7$ - 1.8, which is in good accordance with the experimental observation. This result suggests that the FMV pattern at the triplet circular boundary is more favored than at the doublet circular boundary in terms of the bending energy of the polarization field. To compare the spin correlations in experiment and simulation, the spin correlations $\chi$ associated with small bending energies are plotted as a function of $\Delta/R$ in Fig. \ref{fig3}(e), showing consistency with the experimental observations (Fig. \ref{fig2}(c) and (d)).

The bending energy of the experimentally obtained patterns is worthy of investigation. The bending energy linked to the spin correlation shows that the higher the order, the smaller the bending energy for both doublet and triplet circular boundaries (Fig. S4(a) and (b)). At the triplet circular boundary, unlike the bending energy of the FMV pattern in simulation (Fig. \ref{fig3}(d)), the energy tends to increase with increasing $\Delta/R$. In the experiment, when topological defects with different charges are located very close to each other at the boundary of a small $\Delta/R$, they can annihilate each other, making it easier for a single vortex to form rather than three co-rotating vortices, thus reducing the bending energy.

Macroscopic flows cannot easily be controlled in active turbulence because of intrinsic hydrodynamic instability \cite{ramaswamy2010mechanics,marchetti2013hydrodynamics,aranson2022bacterial}; however, in microscopic individual bacterial bases, chirality exists in the swimming direction. When bacteria swim near the solid surface, they exhibit chiral swimming because the direction of flagellar rotation causes their swimming to follow a unidirectional circular trajectory \cite{frymier1995three,drescher2009dancing,di2011swimming}. We previously found that chiral vortices reflecting swimming chirality appear when confined in microwells with asymmetric interfaces (the top and bottom surfaces were the oil--water interface and PDMS substrate, respectively) (Fig. \ref{fig4}(a)) \cite{beppu2021edge}. The mechanism is explained as follows. The interaction between the chiral motion of the bacteria and the lateral boundary walls results in biased movement at the edge towards the CW and CCW direction above the bottom substrate and beneath the top interface, respectively, when viewed from above (Fig. \ref{fig4}(b)). Unlike the microwells with symmetric interfaces, in the asymmetric system, the density of bacteria at the top interface is approximately 10\% higher than that at the bottom substrate. Hence, this preponderance of the top interface in the competition between opposite chiral motions at the lateral boundary wall leads to a predominant CCW chiral vortex. In mesoscopic turbulent flow patterns, frustrated vortices are responsible for controlling large rotational flows, while, at the microscopic level, chirality ensures the directional control of the flow, leading to larger coherent flows \cite{negi2023geometry,lowen2016chirality}. Next, we investigated the relationship between frustrated vortex pairs and the chirality of vortex rotation.

In a confined microwell with asymmetric interfaces (Fig. \ref{fig4}(a)), for both the doublet and triplet boundaries, the transition point from the FMV to AFMV pattern (Fig. \ref{fig4}(c) and Movie 2) and to frustrated AFMV pattern (Fig. \ref{fig4}(d) and Movie 2) is $\Delta _c/R \approx 1.9$. This implies that chirality is essential in enhancing FMV with geometric frustration. Based on the values of the spin correlation function under two geometric constraints, compared with the chiral vortex with a doublet circular boundary, the chiral-frustrated AFMV maintained a higher spin correlation and remained large up to a region of slightly higher $\Delta/R$ (Fig. \ref{fig4}(e) and (f)). Thus, the interplay between chiral rotation and geometric frustration is effective in directing the collective rotational flow in the broad geometric conditions of a confined space. Since the chiral edge currents are due to the interaction of the bacteria with the lateral boundary, from the standpoint of energy, the chiral effect should act as an effective energy that forces the bacteria to align in one direction (CCW) at the boundary. Given that this energy is increased by the formation of AFMV patterns, it is reasonable that the FMV pattern becomes more favorable and the transition point shifts to a larger $\Delta/R$. A quantitative energy-based study of the pairing order of chiral vortices would be an interesting future direction.

\section{Discussion}

In this study, we conducted an experimental analysis of confined active matter with frustration, using interacting bacterial vortices as a model system. Results showed that no frustration occurred when two vortices interacted and that the transition from the FMV pattern to the AFMV pattern occurred with $\Delta_c/R \approx 1.3$-1.4 \cite{beppu2017geometry,beppu2021edge}. By contrast, the transition from the FMV to the frustrated AFMV pattern with three interacting vortices occurred with $\Delta_c/R \approx 1.7$-1.8 greater than that in the case of two interacting vortices. The theoretical model revealed a shift in the transition point, where the bending energy of the FMV bacterial orientation field was more stable in a triplet circular boundary with a greater $\Delta_c/R$. Our findings indicate that frustration in vortex pairing can alter the geometric dependence of pairing-order transition, highlighting the importance of not only the spacing between interacting vortices but also their geometrical locations.

Exploring the effects of interactions with boundary conditions on the collective motion of active matter can be advantageous for designing microdevices. Irregular patterns can occur in bacterial suspensions with frustrated vortex configurations \cite{wioland2016ferromagnetic,nishiguchi2018engineering}. However, conventional boundary shapes that impose regular geometric confinement can be used to control the flow direction in such configurations \cite{reinken2020organizing}. When addressing irregular boundary geometries, such as porous media, a frustrated vortex pattern can interfere with the distribution of bacterial density \cite{alonso2019transport}. Employing a disordered spacing geometry is an extended challenge for understanding the pattern formation that occurs during interactions with boundaries.

We demonstrated that the shift in the transition point from FMV to frustrated AFMV can be explained by considering the orientation interaction within the triplet circular boundary. This model discusses a stable pattern arising from the orientation bending energy in a scenario where two or three vortices interact at constant spacing. This pertains to the orientation interaction among active matter within the confined space as opposed to the boundary vicinity. We observed that the value of the transition point $\Delta_c/R = 1.2$-1.3 was slightly lower than that of the experimental observation. In our previous studies \cite{beppu2017geometry,beppu2021edge,araki2021controlling}, a Vicsek-style model was used to investigate the transition from FMV to AFMV, assuming that bacteria move along a curved boundary and align their orientation with polar interactions, resulting in $\Delta_c/R = \sqrt{2}$. This model focuses on the alignment of their orientation at the tipping point of the overlapping circular boundary. Therefore, the region in which the orientation interaction was considered was different in the two models, resulting in a small deviation at the transition point. A previous study\cite{liu2021viscoelastic} has reported that a large-scale bacterial vortex can be stabilized in viscoelastic solutions. It is also important to verify with a triplet pair of such large vortices whether the bending energy of the orientations presented in this study makes a significant contribution to frustrated vortex pairing.

Theoretical extensions using the TTSH equation can offer a solution to the unresolved problem of active turbulence. As mentioned previously, the TTSH equation is widely used to investigate the active turbulence observed in bacterial suspensions and explain the AFMV pattern in periodic spacing obstacles \cite{reinken2020organizing,schimming2023vortex}. A recent study \cite{shiratani2023route} has extensively investigated the effect of the geometrical confinement on active turbulence using high-resolution computations of this equation and found the geometry-dependent transition from a single-vortex-like pattern to AFMV under the boundary condition of two overlapping circles. In addition, by considering the slip flow near the boundary, an oscillatory flow reversal can be observed in the edge flow dynamics \cite{matsukiyo2023edge}. However, questions still remain regarding the theoretical model and experimental observations, such as the structural transition of the emerging flow and the stability of the edge flow, which pose a tantalizing challenge in confined active turbulence.

\section*{Acknowledgements}
We thank Y. Sumino for the fruitful discussion. This work was supported by Grant-in-Aid for Scientific Research on Innovative Areas ``Molecular Engines'' (18H05427), Grant-in-Aid for Transformative Research Areas (A) (23H04711, 23H04599), Grant-in-Aid for Scientific Research (B) (20H01872, 23H01144), JSPS Core-to-Core Program “Advanced core-to-core network for the physics of self-organizing active matter" (JPJSCCA20230002), and JST FOREST Grant JPMJFR2239. K.B. acknowledges support from the Overseas Postdoctoral Fellowship of the Uehara Memorial Foundation.

\bibliography{frustration}

\end{document}